\documentclass[%
 aip,
 amsmath,amssymb,
 reprint,%
]{revtex4-1}

\usepackage{graphicx}% Include figure files
\usepackage{dcolumn}% Align table columns on decimal point
\usepackage{bm}% bold math

\usepackage[utf8]{inputenc}
\usepackage[T1]{fontenc}
\usepackage{mathptmx}
\usepackage{etoolbox}
\usepackage{gensymb}
\usepackage[english]{babel}
\usepackage{chemformula}
\usepackage{float}

\makeatletter
\def\@email#1#2{%
 \endgroup
 \patchcmd{\titleblock@produce}
  {\frontmatter@RRAPformat}
  {\frontmatter@RRAPformat{\produce@RRAP{*#1\href{mailto:#2}{#2}}}\frontmatter@RRAPformat}
  {}{}
}%
\makeatother
\begin{document}

\preprint{AIP/123-QED}

\title{Ultrafast laser interaction with transparent multi-layer \ch{SiO2}/\ch{Si3N4} films}

\author{R. Ricca}
 \email{ruben.ricca@epfl.ch}
\affiliation{ 
Galatea Laboratory, IMT/STI, Ecole Polytechnique Fédérale de Lausanne (EPFL), Rue de la Maladière 71b, 2000 Neuchâtel, Switzerland
}%

\author{V. Boureau}
\affiliation{%
Interdisciplinary Centre for Electron Microscopy, École Polytechnique Fédérale de Lausanne (EPFL), 1015 Lausanne, Switzerland}%

\author{Y. Bellouard}%
\affiliation{ 
Galatea Laboratory, IMT/STI, Ecole Polytechnique Fédérale de Lausanne (EPFL), Rue de la Maladière 71b, 2000 Neuchâtel, Switzerland
}%

\date{\today}

\begin{abstract}
We investigate the use of ultrafast lasers exposure to induce localized crystallization and elemental redistribution in amorphous dielectric multi-layers, composed of alternating \ch{Si3N4} and \ch{SiO2} layers of sub-micron thickness. Specifically, we report on the occurrence of a laser-induced elemental intermixing process and on the presence of silicon nanocrystals clusters localized within the multi-layers structure. The spatial distribution of these clusters goes significantly beyond the zone under direct laser exposure providing evidences of energy being channeled transversely to the laser propagation axis at the interface of the nanoscale layers. Thanks to the extreme conditions reigning during laser exposure, this process transposed to various materials may offer a pathway for local and selective crystallization of a variety of compounds and phases, difficult to obtain otherwise.
\end{abstract}

\maketitle

%%%%%%%%%%%%%%%%%%%%%%%%%%  body  %%%%%%%%%%%%%%%%%%%%%%%%%%
\section{Introduction}

The formation of nanocrystalline phases in a dielectric host under intense laser field conditions is motivated by the interest in having nanocrystals dispersed in a transparent and amorphous matrix, in which the nanocrystals are used to locally control electrical and optical properties \cite{lehningerReviewGeNanocrystals2018,liuSemiconductorSolidSolutionNanostructures2017}. In these composites, the dielectric layer not only defines a flat optical interface, but also shields the nanocrystals from the environment, preventing undesired reactions with it. These composites are suitable for photonics \cite{prioloSiliconNanostructuresPhotonics2014}, photovoltaics \cite{conibeerSiliconQuantumDot2008} and memory applications \cite{bonafosSiGeNanocrystals2012}, among others. 

In this context, an intermixing process that consists in forming a new material compound by mixing multiple, stacked layers of various compositions is particularly interesting as it offers a potential means to obtain arbitrary crystal compositions.   

So far, intermixing processes have been reported by performing a temperature annealing (for instance to precisely control the stoichiometry in intermetallics \cite{lehnertNewFabricationProcess1999a} or to form silicon quantum dots in SiC/\ch{SiO_x} hetero-super-lattices \cite{dingSiliconQuantumDot2011a,sirletoEnhancedStimulatedRaman2008}), by ion-beam-induced crystallization \cite{cherkovaLightemittingSiNanostructures2017} and by exposure to excimer lasers radiation \cite{xuElectroluminescenceDevicesBased2014,gontadStudyExcimerLaser2013}. 

\begin{figure}[ht]
\centering\includegraphics[width=0.5\textwidth,height=\textheight,keepaspectratio]{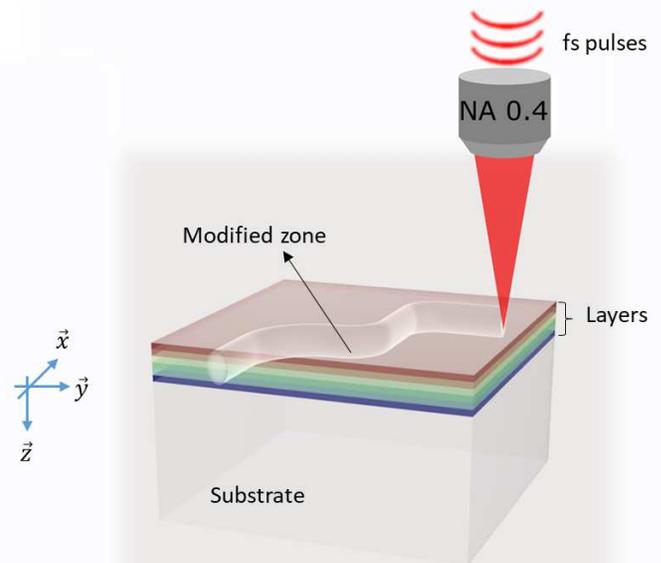}
	\caption{\label{fig-setup}Working principle of multi-layer ultrafast laser modification. A substrate with a stack of thin transparent films (each thinner than the laser-wavelength) is exposed to a femtosecond laser to induce local intermixing and phase transitions in the material, following an arbitrary trajectory.}
\end{figure}

Compared to the reported methods, the use of ultrafast lasers is potentially attractive for multiple reasons. First, as it is based on non-linear absorption processes, ultrafast laser pulses allow the formation of spatially confined modified zones with sizes below the diffraction limit and placed at an arbitrary depth within the layers. Second, as the interaction is seeded by multi-photon processes, any materials transparent to the laser wavelength can be processed. Furthermore, the confined extreme peak intensity (>TW/cm$^2$) opens the prospects for the formation of high-pressure phases as demonstrated in sapphire \cite{juodkazisLaserInducedMicroexplosionConfined2006, gamalyLasermatterInteractionBulk2006}. In addition, some preliminary work on femtosecond laser-induced nano-crystallization has been reported in single layer materials made of silicon oxide and silicon nitride, respectively  \cite{volodinFemtosecondLaserInduced2010, korchaginaCrystallizationAmorphousSi2012} and, despite some degree of surface ablation, in semi-conductor multi-layers systems composed of amorphous Si (a-Si) and germanium (a-Ge) layers \cite{kolchinFemtosecondLaserAnnealing2020}. In the latter, although Raman analysis was performed to identify the presence of crystalline Ge (c-Ge), no detailed analysis of the crystallites and related formation mechanism was proposed. 

Femtosecond laser-dielectric multilayers interaction involves unusual modes of propagation and reflection of electromagnetic waves due to the layered stack itself. The periodic arrangement of alternating layers of different materials can be considered as a one-dimensional photonic crystal, whose dielectric function only varies along the vertical axis \cite{joannopoulosPhotonicCrystalsMoldingthe2007}. In the case of femtosecond laser exposure, the differences in individual layers' non-linear responses involve a higher level of complexity. Ionization can for instance take place in discrete, non-contiguous locations along the vertical axis. Csajbok \textit{et al.} \cite{csajbokFemtosecondDamageResistance2016,mangoteFemtosecondLaserDamage2012} reported that the intensity of the laser’s electric field along the multilayer stack is periodically modulated along the propagation axis, according to the stack geometry and composition, presenting intensity minima and maxima. As a result, in standard designs of distributed Bragg reflectors (DBRs), the multilayers weakest points with respect to the damage threshold reside at the interfaces with the high-index materials, so that the materials resistance to the laser can be improved by modifying the layers design, shifting the electric field’s intensity peaks towards the lower-index layers \cite{shunlichenEffectStandingwaveField2011,angelovOpticalBreakdownMultilayer2013,schiltzModificationMultilayerMirror2014,lifengAnalysisLaserinducedTransient2016, chorelRobustOptimizationLaser2018}. The intensity has also been found to be enhanced locally by pits, scratches and more generally defects, resulting in higher damages of the multilayers in these regions \cite{chaiLaserresistanceSensitivitySubstrate2016,kozlovMechanismsPicosecondLaserinduced2019a,gallaisInfluenceNodularDefects2014,wolfeFabricationMitigationPits2011}. Investigations on the role of laser and materials parameters have shown a link between refractive indices and geometrical parameters (thickness and number of layers) and inter-layer stress-states after exposure, with the intensity profile of the electric field defining the dynamics and nature of the materials modifications  \cite{gallaisTransientInterferenceImplications2010,longFabricationDamageCharacteristics2019a,lifengThermalstressAccumulationAntireflective2015,sunDynamicsFemtosecondLaser2006,kumarQuantizedStructuringTransparent2014}. 

Building-up on these previous works, this paper aims at exploring ultrafast laser-matter interaction with layered dielectrics, specifically from the point of view of intermixing and nanocrystalline phases formation. Using advanced material characterization tools, we systematically investigate the morphology of laser affected zones and the location of the nanocrystallites as well as their crystallographic structures.

\section{Materials and methods}
\begin{figure}[ht]
\centering\includegraphics[width=0.5\textwidth,height=\textheight,keepaspectratio]{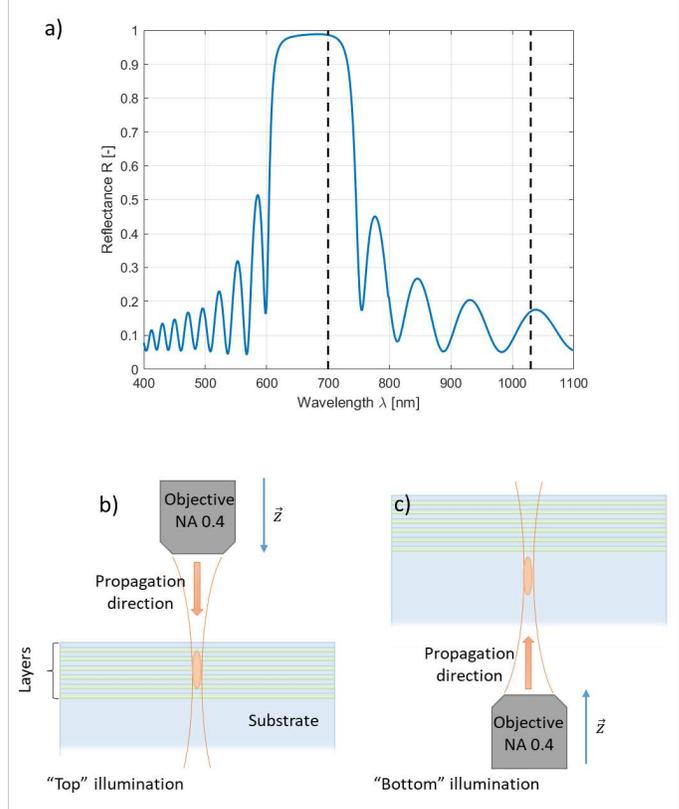}
	\caption{\label{experimental}a) Reflectance spectra of the samples. 700 and 1030 nm wavelengths positions are marked with black dashed lines. b) and c) illustrate the two exposure modalities explored in this work: the "top" illumination and the "bottom" illumination (through the substrate) of the layer stack, respectively.}
\end{figure}

As a case study, we consider a multilayer system consisting of alternating silicon oxide (\ch{SiO2}) and silicon nitride (\ch{Si3N4}) layers deposited on a fused silica substrates by plasma-enhanced chemical vapor deposition (PECVD). The individual layer thicknesses are 119 nm (\ch{SiO2}) and 87~nm (\ch{Si3N4}), respectively, and a total of 13 double layers were deposited. Samples with similar designs are for instance used in Bloch wave propagation experiments \cite{yuManipulatingBlochSurface2014}. The multilayer stack parameters were chosen to maximize the reflectance at normal incidence and according to a standard quarter wave-design (Fig.~\ref{experimental}a). Specifically, it has a high transmittance at the incident laser wavelength (1030 nm) and a reflectance peak around 700 nm. This particular reflectance wavelength range corresponds to a maximum in sensitivity for standard imagers and to the wavelength range of the laser, once travelling through the silica substrate in the case of backside illumination. The deposition process was done using an Oxford PlasmaLab 80. The deposited multilayer stack is originally amorphous in nature, as confirmed by Raman spectroscopy (Fig.~\ref{fig-cryst-conf}).

The ultrafast laser source (Yuzu from Amplitude Systèmes) delivers pulses of 270 fs at a central wavelength of 1030 nm and at a repetition rate of 1 MHz. The optical setup consists in a mirror relay from source to sample, with an automated rotating half-wave plate and a polarized beam splitter for power control and a half-wave plate before the objective to linearly adjust the polarization (perpendicular, parallel, or in between the two states). The beam is focused in the specimen using a 20x objective lens with a numerical aperture (NA) of 0.4, which can be moved along the vertical direction to position the focal spot at various heights inside and outside the sample volume. A motorized stage (Micos, Ultra-HR) controls the movement of the sample in the horizontal plane. Displacement resolution along each axis is about 100 nm, with a repeatability around 200 nm. Two laser exposure strategies are investigated. A first one, referred to as "top" illumination, consists of focusing the laser directly on the multilayer stack (see Fig.~\ref{experimental}b). The second one, referred to as "bottom" illumination, has the beam passing through the substrate first and focused a few microns below the layer's interface (Fig.~\ref{experimental}c). The purpose of this second exposure approach is to study the effect of the pulse propagation through the bulk substrate (before reaching the focal spot) on the modifications dynamics. In both cases, the laser beam is scanned over the specimen at constant speed. We define the cumulative exposure as the number \textit{N} of subsequent pulses emitted at a frequency \textit{f} and translated at a velocity \textit{v}, impacting the area defined by the beam waist $w_0$, such as $N= w_0 f/v$.

On average, between 450 and 900 pulses impact a same zone, depending on the repetition rate used (250 or 500 kHz), and considering a scan speed of 1 mm s$^{-1}$, meaning that cumulative exposure is achieved. 

To compare exposures conditions generated with different laser parameters, we use here a quantity defined as deposited energy $E_d$ (or net fluence), expressed in J/cm$^2$, which accounts for the exposure dose and is defined in Eq.(\ref{equationE}) \cite{rajeshFastFemtosecondLaser2010}:

\begin{equation}
	E_d=\frac{4E_p f}{\pi\omega v}
	\label{equationE}
\end{equation}

Where $E_p$ is the pulse energy, $f$ the laser repetition rate, $v$ the scan speed and $\omega$ the beam waist at the focal spot. The beam diameter at focus is 1.8~$\mu$m, confirmed both by beam propagation simulations and actual beam diameter measurement (see supplemental material in Torun \textit{et al.} \cite{torunDirectwriteLaserinducedSelforganization2021}). The corresponding Rayleigh length of 2.47~$\mu$m (taking into account the quality factor of the beam, or M$^2$ value, of the laser) is an order of magnitude greater than the height positioning resolution of the stage (<0.1 $\mu$m), which guarantees an accurate positioning of the beam on and within the specimen for the confocal parameters and the specimen characteristics dimensions considered in this study. The energy density will also be expressed with respect to individual pulse fluence, $F=4 E_p / \pi \omega^2$, in J cm$^{-2}$, to differentiate the role of the field-strength in laser-induced structural modifications, and of peak irradiance $P_p=0.94 E_p/(\tau_p \pi \omega^2)$ to include the pulse duration parameter ($\tau_p$).

The changes in microstructure are studied using Raman spectroscopy, scanning electron microscopy (SEM) and transmission electron microscopy (TEM), for which specimen preparation is done with a focused ion beam (FIB). Raman spectroscopy is performed with a Horiba Jobin-Yvon LabRam HR instrument, using a laser excitation wavelength of 532 nm focused with a 0.9 NA objective. The excitation laser average power is limited to 4 mW to limit local heating of the specimen during the Raman analysis that might alter the laser-induced nanostructures under observation. SEM observations and cross-sections cuts are done on a FEI Nova 600 NanoLab dual-beam SEM/FIB, running at 5/30 kV respectively. Samples preparation for TEM imaging is realized  with a Zeiss NVision 40 dual-beam SEM/FIB, to produce lamellas from specimen cross-sections thinned down to thicknesses of about 100 nm. TEM observations are done using a Talos F200S from ThermoFisher operating at 200~kV.

\section{Experimental results}
\subsection{Evidence of modifications and crystallization}
Exposing regions to both top and bottom illumination conditions leads to different structural morphological changes depending on the exposure parameters, including an area under compression on the substrate, localized voids in the layers, evidences of intermixing between layers, partial ejection of particles in the surrounding of the modified zones and partial to full ablation of the layers. 

\begin{figure}[ht]
\centering\includegraphics[width=0.5\textwidth,height=\textheight,keepaspectratio]{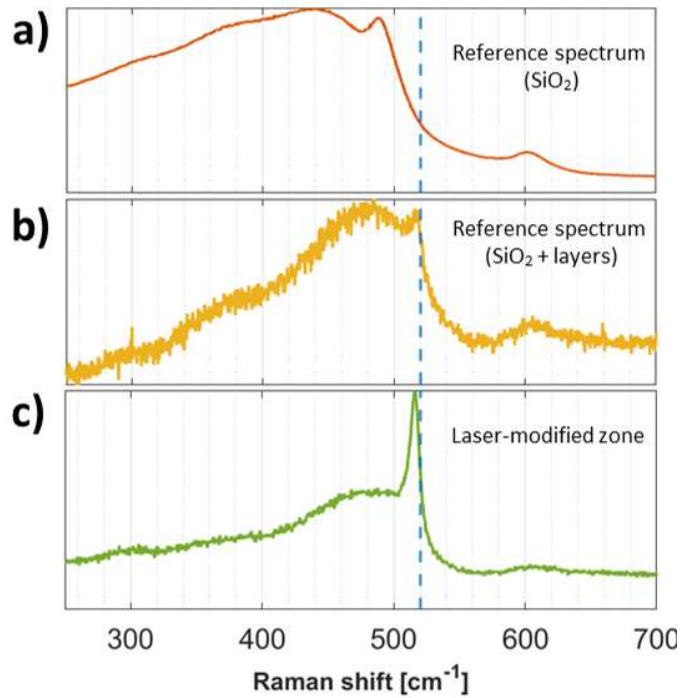}
	\caption{\label{fig-cryst-conf}Raman spectra of un-exposed and exposed regions. a) reference curve for the amorphous silica \ch{SiO2} substrate (measured on the back side of the samples, far away from the multi-layers), b)  pristine multi-layers surface, and c) measured spectrum from a laser-affected zone. The position of crystalline silicon peak (520 cm$^{-1}$) \cite{parkerRamanScatteringSilicon1967} is shown with a dotted blue line. Note that due to a lower Raman laser power used to preserve features of interest in the specimen, the spectra related to multi-layers regions are noisier.}
\end{figure}

We first performed Raman observations on modified specimens. The Raman spectra of the bare substrate, the covered substrate with multi-layers and of laser-affected zone are shown in (Fig.~\ref{fig-cryst-conf}). A noticeable shift and narrowing of the main band is observed after deposition of the multi-layers (Fig.~\ref{fig-cryst-conf}b). We attribute these effects to the presence of residual deposition stress between  multi-layers and between the stack of layers and the support substrate, itself made of silica. Near and within laser-affected zones (LAZ), the presence of a narrow and intense peak, centered around 515 cm$^{-1}$ (Fig.~\ref{fig-cryst-conf}c) indicates a localized crystallization event that took place within the laser affected zone, which can be attributed to a crystalline silicon phase (c-Si) \cite{parkerRamanScatteringSilicon1967}. 

\begin{figure*}[ht!]
\centering\includegraphics[width=0.95\textwidth,height=\textheight,keepaspectratio]{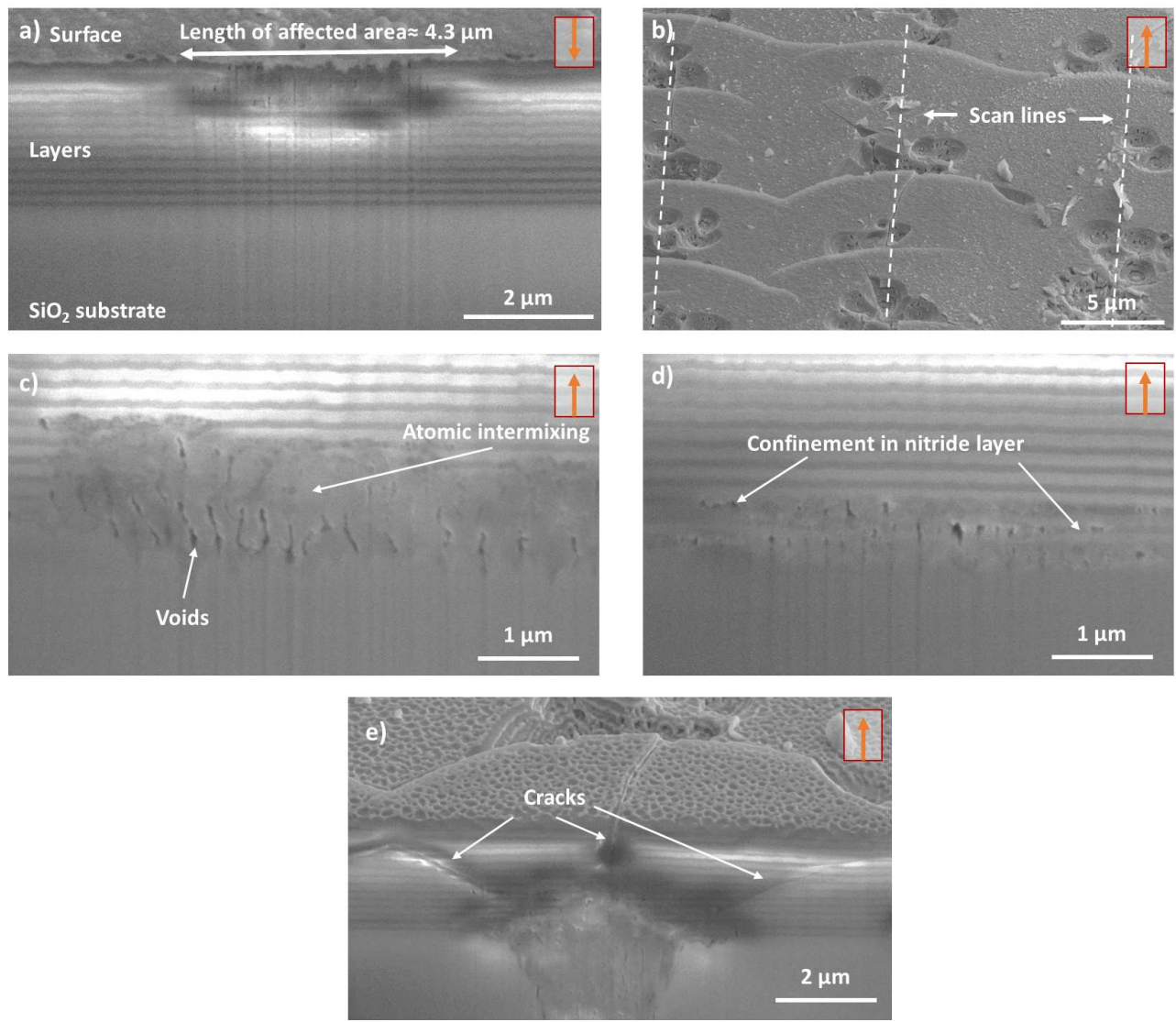}
	\caption{\label{fig-SEM-A}SEM images of laser-exposed regions. In a) and c)-e), a FIB milling was performed to expose the cross-sections. The propagation direction of the laser beam is shown with an orange arrow on the upper right of each image. a) top illumination of the multi-layer shows a section with greater intermixing of the layers and large affected area, with respect to the beam size. b) specimen surface in the case of a bottom illumination, showing localized fractured pits and longitudinal cracks, transversely to scan lines directions. c-e) modified zones in bottom illumination mode, at a repetition rate of 100 kHz and scan speed of 1 mm s$^{-1}$. The pulse energies are of 500 nJ in c), 250 nJ in d) and e) (deposited energies 7.07 and 3.54 kJ cm$^{-2}$, pulse fluence 78.6 and 39.3 J cm$^{-2}$, peak irradiance of 68.4 and 34.2 TW cm$^{-2}$, respectively) while the distance between layers and focal spot is 40 $\mu$m in c) and d), 10 $\mu$m in e).}
\end{figure*}

A small, but persistent shift in the interpreted Si peak position compared to the reported c-Si values (5-10 cm$^{-1}$ on average, up to 20 cm$^{-1}$) is observed (data not shown here), and is attributed to residual stress induced by the lattice mismatch between crystalline regions and the surrounding matrix \cite{kangApplicationRamanSpectroscopy2005}. More specifically, sharper peaks are shifted towards wavenumbers higher than 515 cm$^{-1}$, while the shallower peaks (which indicate a lighter presence of crystallites) are shifted towards lower wavenumbers.
Cross-sections of laser affected zones were observed using a SEM to examine the post-exposure morphology of the multilayers, both in top (Fig.~\ref{fig-SEM-A}a) and bottom (Fig.~\ref{fig-SEM-A}c-e) exposure configurations. Prominent features are visible, depending on the exposure conditions, including evidences of partial ablations (see surface in Fig.~\ref{fig-SEM-A}b), formation of periodic micro-voids (see cross-section in Fig.~\ref{fig-SEM-A}c, d), intermixing between layers (Fig.~\ref{fig-SEM-A}a, d, e), cracks across layers (Fig.~\ref{fig-SEM-A}e) as well as  damages at the layer-substrate interface (Fig.~\ref{fig-SEM-A}c-e). Fig.~\ref{fig-SEM-A}d additionally shows a confinement of the modifications in the nitride layers, suggesting that non-linear absorption is initiated in \ch{Si3N4} at a lower energy threshold than \ch{SiO2}. Raman spectroscopy highlights the presence of crystalline phases in most of these cases (barred the most violent instances of ablation).
Let us examine further the cases of bottom illumination shown in Fig.~\ref{fig-SEM-A}c), d) and e). The effective wavelength of the incoming laser beam travelling in the substrate is approximately 710 nm, which is within the peak reflectance band of the Bragg reflector. As a result, the incoming light is reflected at the interface. The first layers of the Bragg mirror are subjected to the highest energy and consequently are the first to be affected, which is confirmed in c) and d). In the case of e), the ionization of the substrate overlaps with the one of the Bragg mirror first layers. In here, the rapid plasma expansion may explain the formation of cracks distributed symmetrically and normally to the shock front expected to propagate from the Bragg mirror-substrate interface. Higher pulse energies and shorter layer-focal spot distances are expected to enhance the ionization rate and result in broader modifications, as visible in e). 

\subsection{Intermixing and nano-crystalline phases}
\begin{figure*}[ht!]
\centering\includegraphics[width=0.855\textwidth,height=\textheight,keepaspectratio]{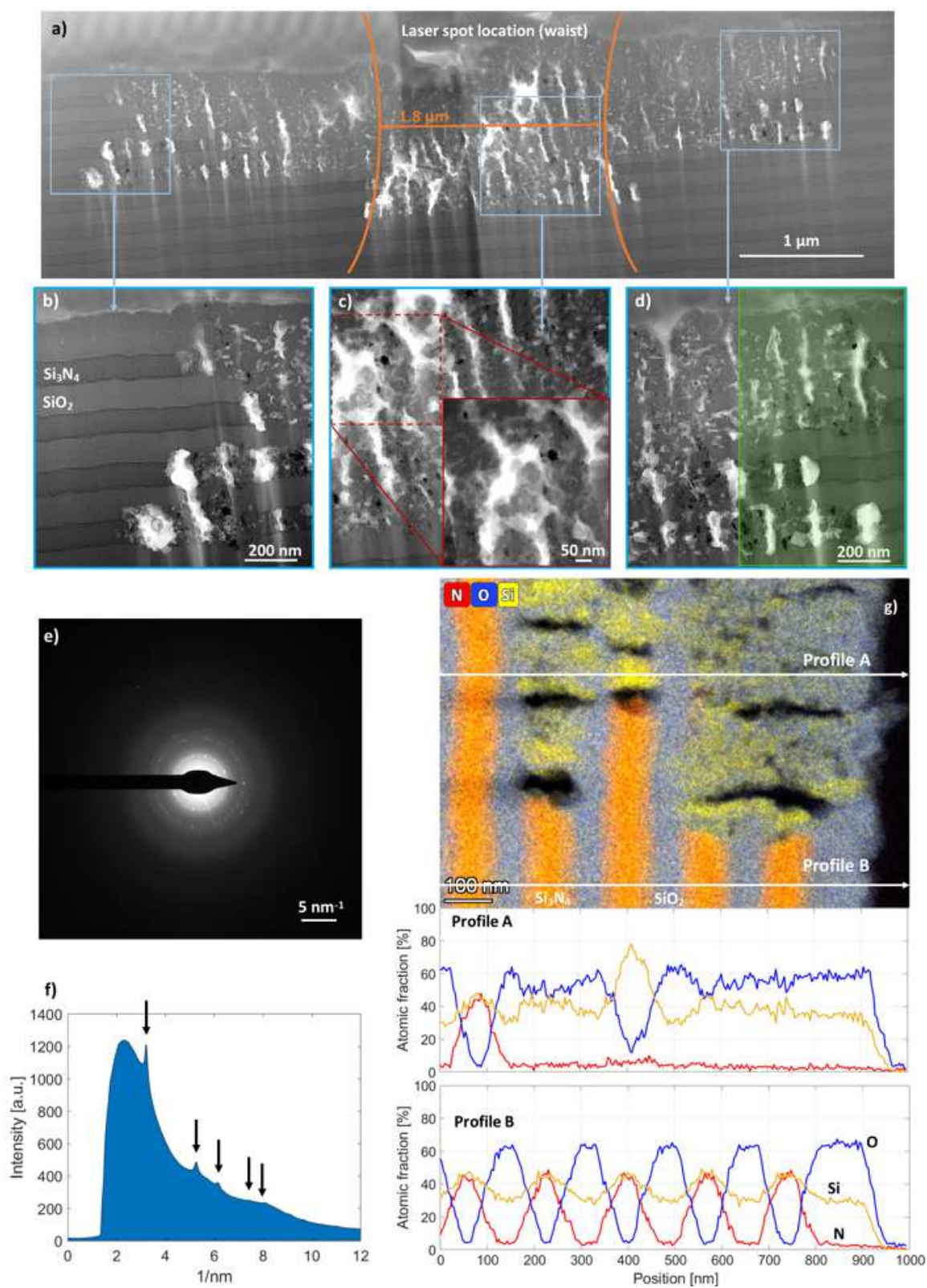}
	\caption{\label{TEM_1} a) BF TEM image of the sample exposed in top illumination configuration, corresponding to a cross section of an exposed line. b)-d) Higher magnification BF TEM images of left, central and right parts of the damaged region (indicated by blue squares, in a). e) SAED pattern of a 850 nm-diameter region of the damaged region and f) rotational averaging of the SAED showing the positions of the crystal peaks over the diffuse background. g) EDS elemental color-map (net counts) of the right part of the damaged region, indicated in green on image d) and rotated 90$^\circ$ clockwise, from where two profiles of atomic fractions are extracted with an integration width of 35 nm.}
\end{figure*}

TEM observations of cross-sections of the exposed regions were used to localize crystalline zones within the laser-affected zones. A first observation was done on a sample exposed in the "top illumination" configuration, at a repetition rate of 250 kHz, scan speed of 1 mm s$^{-1}$, a deposited energy of 2.49 kJ cm$^{-2}$, a pulse fluence of 5.54 J cm$^{-2}$ and a peak irradiance of 4.8 TW cm$^{-2}$. A bright-field (BF) TEM image of the sample is shown in Fig.~\ref{TEM_1}a. The various layers are clearly visible, with the \ch{Si3N4} layers (darker) alternating with the \ch{SiO2} layers (lighter). Empty spaces appear white, while nano-crystalline regions appear as small dark regions, when oriented in strong diffracting conditions. At a first glance, the presence of nano-crystalline regions observed in Raman measurements is confirmed by the presence of isolated dark clusters in the BF images. The dimensions of these nano-crystallites vary significantly, but seem not to exceed 50 nm (Fig.~\ref{TEM_1}b-d). 

Interestingly, the dimensions of the damaged area exceed several times the dimensions of the laser beam waist at the focal point, suggesting that part of the incoming laser energy was propagated away from the zone under direct laser exposure through the denser layers. The location of the beam waist is highlighted in orange in Fig.~\ref{TEM_1}a, and is assumed to be centered on the modified zone due to its symmetry. This central zone experienced higher intensity and corresponds to a zone where the deepest modifications are observed. The average width of this modified central portion is 2~$\pm$~0.25 $\mu$m), which is in good agreement with the measured spot waist at the focus. Remarkably, crystallites are found as far as 2 $\mu$m away from the edge of this central section corresponding to the beam waist. These nano-crystalline phases observed with the TEM are assumed to be at the origin of the crystalline peak found in the Raman spectra (see Fig.~\ref{fig-cryst-conf}).
Other salient TEM observations are the sharp boundaries, both along and transverse to the laser propagation direction, that define the edges of the damaged zones. Here, modifications are confined in the higher index layers (\ch{Si3N4}), and stop abruptly in the transverse direction a few microns away from the focal waist. This lack of gradual transitions is also observed along the propagation direction. Note that the non-linear nature of the interaction is illustrated by the abrupt transition between modified/non-modified \ch{Si3N4} layers. Although based on post-exposure investigations, this observation brings evidences that part of the incoming laser-beam energy was channeled away over significant distances from the beam waist through the higher index \ch{Si3N4} layers. This happened in a non-thermal regime as the modification remained mostly confined in the \ch{Si3N4} nano-layers, since the boundaries with the \ch{SiO2} layers are still distinctly visible. 
 This observation suggests an ionization mechanism driven by evanescent-like and coupled-waves propagating in the \ch{Si3N4} higher-density layer. The initiation and partial confinement of the modifications in the \ch{Si3N4} layers is consistent with the properties of quarter-wave design Bragg reflector, for which the pulse's electric field intensity is modulated in such a way to have maxima preferentially located in the higher-index layers \cite{csajbokFemtosecondDamageResistance2016,mangoteFemtosecondLaserDamage2012}, therefore increasing the probability of localized ionization to take place. Taking into account the lower band-gap of \ch{Si3N4} layers and the Bragg reflector properties, damages are expected to be initiated first in the \ch{Si3N4} layers, which is confirmed by the TEM observations. Furthermore, as seen in figures \ref{TEM_1}a-d, the void regions are distributed periodically and are aligned along the vertical direction (i.e. beam propagation direction), extending along the length of the damaged area.

The crystallographic structure of crystallites found in the central region was investigated by selected area electron diffraction (SAED) (Fig.~\ref{TEM_1}e). A ring pattern is observed due to the random orientation of the crystal grains. The spots are well organized in the reciprocal space, with positions measured from Fig.~\ref{TEM_1}f. The comparison of the peaks distribution with a database hints at a diamond-structured silicon phase, with no other ordered phases detectable (experimental and theoretical values agree within a maximum deviation of 1.4\%, see Table \ref{table-EDX}).

\begin{table}[h!]
\caption{\label{table-EDX}%
Interplanar distances measured by SAED compared to the theoretical distances for diamond-Si, in nm$^{-1}$.
}
\centering\begin{tabular}{ccc}
\hline
\textrm{Corresponding plane}&
\textrm{Measured}&
\textrm{Theoretical}\\
\hline
(111) & 3.21 & 3.189\\
(220) & 5.28 & 5.208\\
(113) & 6.15 & 6.107\\
(004) & 7.44 & 7.365\\
(133) & 8.08 & 8.026\\
\hline
\end{tabular}
\end{table}

To investigate the intermixing process, elemental mappings were performed by scanning TEM (STEM) energy-dispersive X-ray spectroscopy (EDS) analysis. The accuracy of the atomic fraction profiles can be estimated using profile B, plotted from the undamaged region in  Fig.~\ref{TEM_1}g. For \ch{SiO2} an atomic percentage of 34\% and 66\%, is measured for silicon and oxygen atoms, respectively, which corresponds to the nominal stoichiometry of \ch{SiO2}. For \ch{Si3N4}, the ratio between elements is more balanced with approximately 50\% of both elements. The nitride layers are slightly more rich in silicon than the stoichiometric compound. Following profile A (Fig.~\ref{TEM_1}g), we notice that in the damaged zones the nitrogen, where initially present, disappeared, being substituted both by oxygen and silicon in the locations where there are no voids. 

Silicon clusters regions, with higher fractions of Si compared to the initial stoichiometry of \ch{SiO2} and \ch{Si3N4}, are clearly identifiable and distributed along the laser-modified zones, inside and outside the optical beam waist. There, the concentration of Si atoms reaches up to 80\%. Note that pure Si regions are not observed in EDS maps due to projection effects of the TEM measurements over the lamella's thickness, which is thicker than the Si clusters. Generally, oxygen is present in a relatively homogeneous concentration, just slightly lower than the stoichiometric \ch{SiO2}, most probably due to the same TEM projection effects. In conclusion, SAED measurements allow to assess the presence of diamond Si nano-crystallites, randomly oriented, while EDS measurements allow to map the distribution of these Si crystallites in the vicinity of the damaged zone, which spans over a wider region than the laser-beam waist. Furthermore, SAED information is supported by Raman results, while EDS mapping of Si crystallites shows a good agreement with the dark contrasts observed in the BF TEM images.

\begin{figure*}[ht]
\centering\includegraphics[width=1\textwidth,height=\textheight,keepaspectratio]{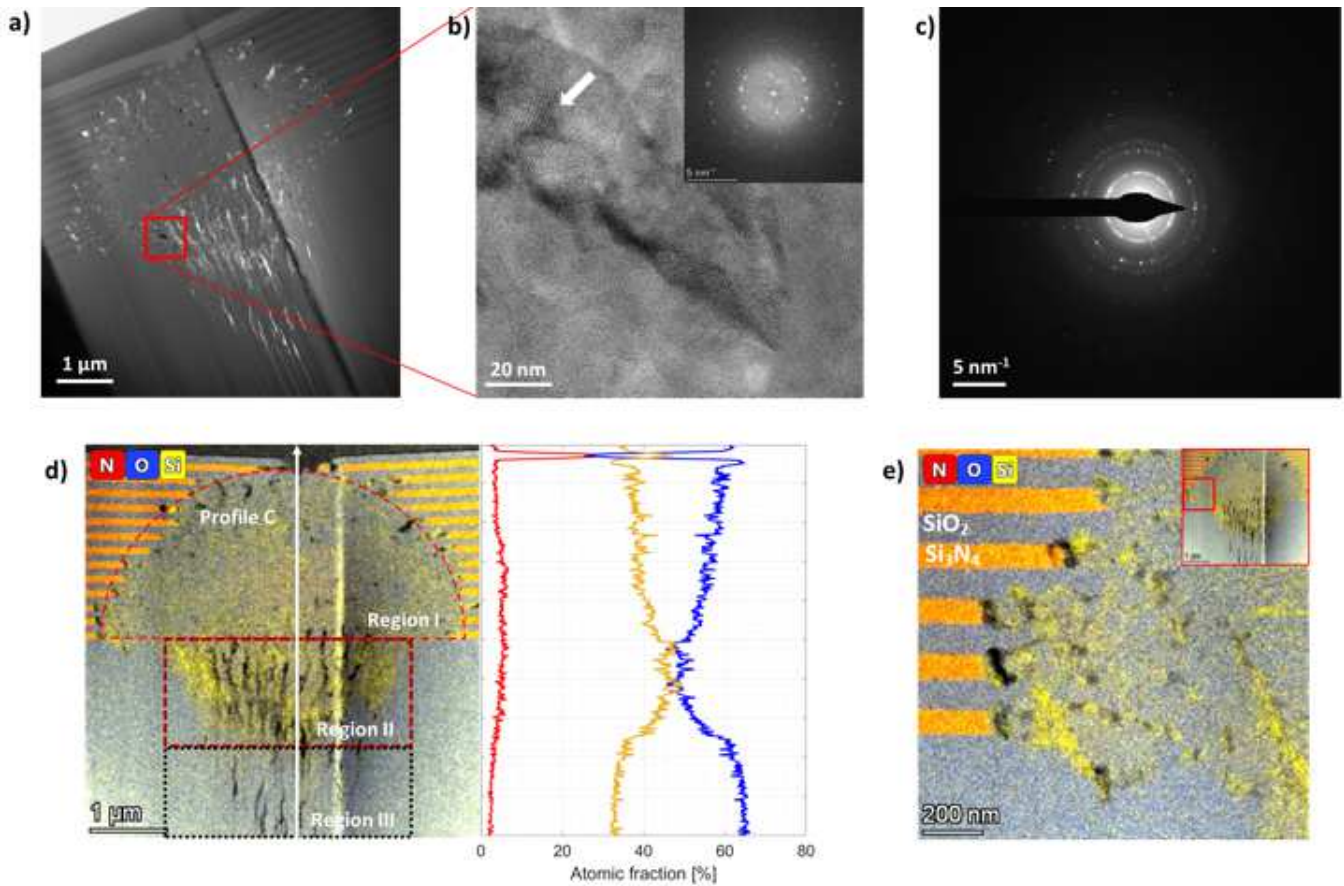}
\caption{\label{TEM_2} a) BF TEM image of the sample after bottom illumination exposure, corresponding to a cross section of an exposed line. b) shows a high-resolution TEM image of a crystallite from the red area, with the Fourier transform of the image in inset that evidences the presence of the crystal planes. c) SAED pattern from a 850 nm-diameter area around the same area. d) EDS elemental color-map (net counts) of the whole damaged region, divided into three distinct zones (I to III) and vertical profile of atomic concentrations in the center of the modified area (along the white arrow), calculated with an integration width of 2 $\mu m$. e) higher magnification map from d).}
\end{figure*}

In the case of bottom illumination (Fig.~\ref{experimental}c), depending on laser parameters and the location of the waist with respect to the layers, three types of modifications are induced: local damages in the form of voids and intermixed regions, internally confined explosion, and layer ablation, respectively. Here, we focus our attention to the case where no ablation occurred (Fig.~\ref{TEM_2}). 

This specific case corresponds to an irradiation with laser pulses fired at a repetition rate of 500 kHz, a scan speed of 1 mm s$^{-1}$ and a pulse energy of 150 nJ, resulting in a deposited energy level of 10.61 kJ cm$^{-2}$, a pulse fluence of 23.58 J cm$^{-2}$ and a peak irradiance of 20.5 TW cm$^{-2}$. The focal point is localized approximately 7.5 $\mu m$ from the substrate/layers interface. 

Like in the previous exposure conditions, crystallites appearing as dark clusters on TEM images (Fig.~\ref{TEM_2}a) are noticed, and are spread around the damaged area. Just like for top illumination, the region affected by the laser is several times wider than the actual beam waist. Fig.~\ref{TEM_2}b shows a high-resolution TEM image of a large crystallite inside the amorphous matrix, which shows Si lattice planes, as evidenced in the inset that shows the Fourier transform. The SAED analysis in multiple locations of the cross-section, as shown in Fig.~\ref{TEM_2}c, confirms that the crystallites are diamond-lattice silicon, showing the same reflections as the one analyzed in Table \ref{table-EDX}.

As for top illumination, the modified area is almost fully depleted of nitrogen (Fig.~\ref{TEM_2}d-e). On one hand, a nitrogen atomic content of 4\% is measured on average in the upper 1~$\mu$m of region II, increasing to 5\% deeper in region II and in the top of region I (see profile of Fig.~\ref{TEM_2}d). Detection of nitrogen within the modified region of the substrate indicates a migration of this element from the  \ch{Si3N4} layers into the substrate during laser exposure. On the other hand, the concentration of Si increases in the damaged areas (see profile of Fig.~\ref{TEM_2}d). Although it was originally composed by stoichiometric amorphous \ch{SiO2} (atomic O/Si ratio of 2), like also measured in the lower part of region III, average atomic ratios of O/Si of 1.5 and 1.1 are measured in region I and in the top of region II, respectively. 

The characteristic dome shape of region I, symmetric with respect to the optical axis, and the somewhat homogeneous mixing of atomic species in its volume, suggest that the Bragg mirror gradually ruptured as ionization of the substrate started. This could lead the broadband radiation scattered from the interface to propagate homogeneously within the dome volume without a preferential direction. 

\subsection{Particle ejection and residues analysis}
In both cases described in the previous section, ejection of particles is observed at high pulse peak fluence and deposited energies. Although the presence of these residues signals a transition towards an ablative regime and as such is of less practical interest, their compositions offer further insights on the modification dynamics. The ejecta concentration logically depends on the deposited energy and, in the bottom-illumination case, on the focal point position, i.e. the closer the focusing to the substrate-multi-layer interfaces, the higher the residues concentration in the ejecta. 

\begin{figure*}[ht!]
\centering\includegraphics[width=0.8\textwidth,height=\textheight,keepaspectratio]{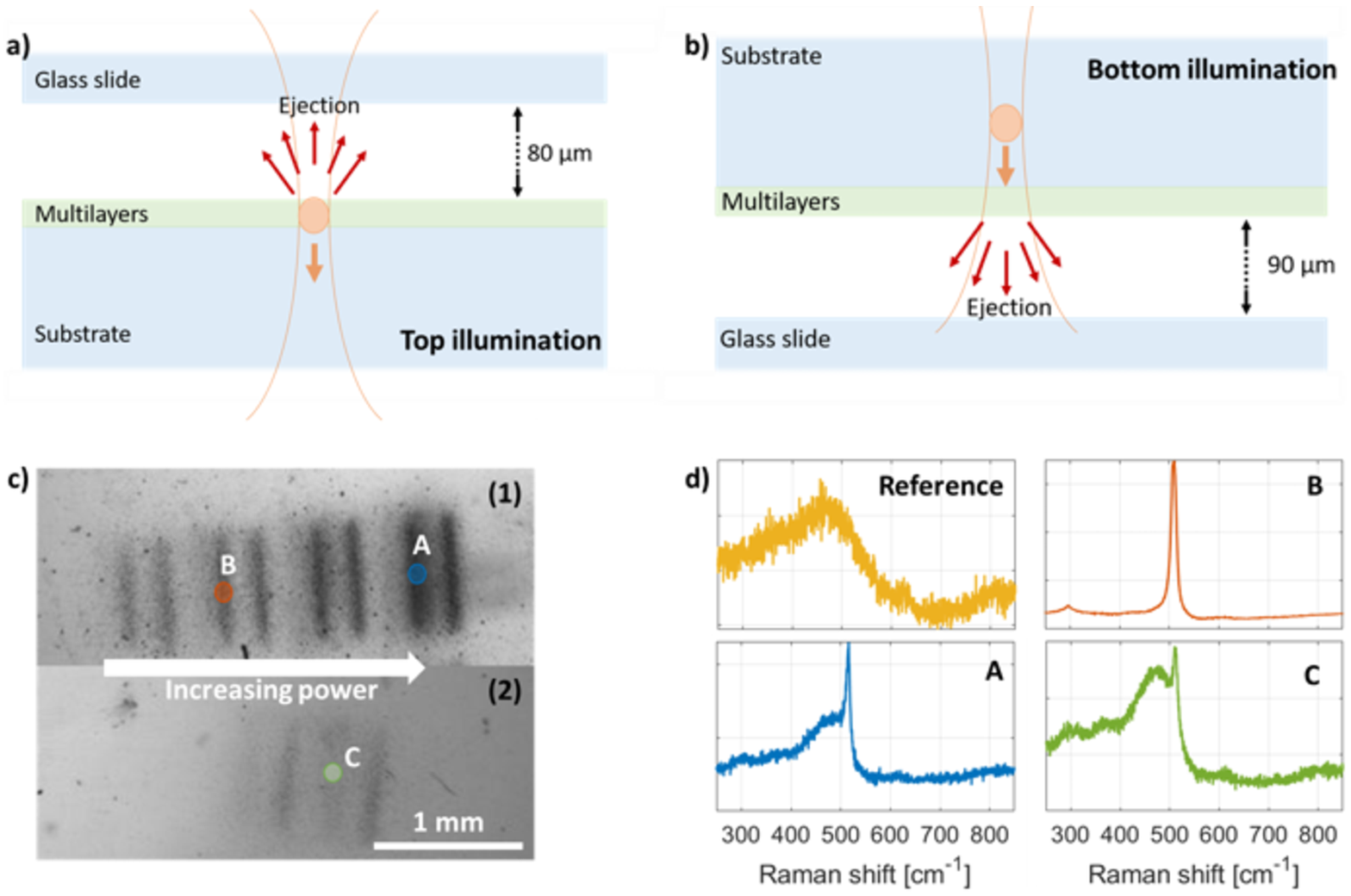}
	\caption{\label{fig-ejection}a-b) configurations of the laser exposure, with the direction of ejection of ablated residues, in the top and bottom exposure cases, respectively. The optical microscope image in c) shows the accumulation of ablated particles on the glass sample holder after bottom illumination (c1), showing an increase in particles concentration for focusing point closer to the multi-layers location, and after top illumination (c2). The colored circles and letters refer to the Raman spectra shown next to it in d). The yellow spectrum is provided as reference and corresponds to the pristine glass slide collector substrate.}
\end{figure*}

These ejected particles are collected on a collecting substrate (here a borosilicate microscope slide), located at a fixed distance from the multi-layers and within the laser beam path (Fig.~\ref{fig-ejection}a-b). Note that as the laser beam is unfocused at the collecting substrate, and hence, the density of energy is not sufficient to trigger non-linear absorption, no modification of the substrate occurs.  The collected material (Fig.~\ref{fig-ejection}c) is subsequently analyzed with Raman spectroscopy by scanning the Raman probe on the surface with a 3 $\mu$m-spacing between measurement points (Fig.~\ref{fig-ejection}d).

A selection of the Raman spectra is shown in Fig.~\ref{fig-ejection}d. These curves are chosen to highlight various spectral occurrences  qualitatively, but do not contain quantitative information related to the crystallites distribution. The residues Raman spectra exhibit crystalline peaks positioned around 515 cm$^{-1}$, consistent with our previous observation of non-ablated exposed regions (Fig.~\ref{fig-cryst-conf}). Further referring to the TEM analysis (Fig.~\ref{TEM_1} and Fig.~\ref{TEM_2}), it is likely that the ejecta contains crystallites of the same phase identified in the non-ablated cases described in the previous paragraphs. These results confirm that both top and bottom exposure conditions lead to similar crystallization conditions. The presence of crystalline clusters in the ejecta logically indicates that the crystallization took place in the early steps of the interaction, before ablation effectively took place. As these experiments are not done with single pulse, crystallization may have gradually occurred, before the particles are propelled outside of the modified zone.  

This experiment is somewhat similar to a Laser-Induced Forward Transfer (LIFT) process, in which ultrafast laser pulses are used to eject material from a selected thin film, effectively resulting in a controlled and localized deposition \cite{serraLaserInducedForwardTransfer2019}. Previous work in this field includes the observation of the deposition of crystalline particles starting from an amorphous source \cite{toetLaserassistedTransferSilicon1999,narazakiNanoMicrodotArray2008} as well as the deposition of transparent layers \cite{banksInfluenceOpticalStanding2009}. Here, starting from controlled multi-layers compositions and thicknesses, the ejection process in our case can potentially be exploited to deposit particles with exotic compositions, blending compounds from the various nanolayers.

\section{Discussion}
Before proposing a phenomenological interpretation of the events unfolding in these experiments, let us first summarize the main results from our observations. Note that these are based on postmortem analysis and, as such, do not provide indications on the exact sequence of events leading to the end-product.

In both exposure configurations, strong intermixing of the atomic species of the layers occurred within the modified zones, and the formation of a silicon nano-crystalline Si phase is observed. A prominent feature is that damaged areas are significantly exceeding the actual beam waist size in focus with no evidence of delamination between nano-layers. Top illumination displays a discrete damage morphology propagating far-off the laser-exposure site within the \ch{Si3N4} layers and defining sharp boundaries between damaged layers, particularly visible at the tail of the modified zone, with no evidence of diffusion of one layer into the adjacent one, suggesting localized energy confinement in the films with fast dynamics.  The extent of these modifications and their confinements to the high-index layers suggest the occurrence of a coupled/evanescent wave propagating sideways with sufficient intensity to cause localized ionization and plasma formation. The nature of this coupled-side wave can be multiple and is discussed further below.  

Bottom illumination yields a structure considerably different than for top illumination. The most noticeable difference is the prevalence of homogeneous atomic intermixing in the modified zones, instead than discretized modifications confined in particular layers, as is mostly the case in the top illumination approach. We label "region II" and "region III", in Fig.~\ref{TEM_2}d), the modified regions localized within the substrate. The two regions are considered separately because region II has a particularly high fraction of Si with respect to the surrounding substrate, while region III is in line with the stoichiometric ratio. In both regions, periodic self-organized voids forming nanoplanes are observed, stopping at a rather well-defined boundary where the interface between the first nano-layer and the substrate was previously found. Above it, a somewhat homogeneous (in sharp contrast with the previous case) dome-shaped area of several microns in diameter is found (labeled "region I" in Fig.~\ref{TEM_2}d) that corresponds to a region with strong atomic intermixing. The dome-shaped structure spreads over a region significantly bigger than the laser waist. As we do not see evidence of thermal dissociation of the material, the plasma was likely confined in a region that is now defining the regions I to III in the end-product. We also note that nitrogen diffused into the substrate in a rather homogeneous manner (see Fig.~\ref{TEM_2}h), which may indicate that the plasma expanded towards the bulk. Despite the fact that the interface between substrate and the nano-layers did vanish (interface between region I and II), the two regions are still clearly distinguishable and, given the presence of N in the substrate, this suggests that region I formed first and regions II ad III after. Region I may have formed first, as the multi-layers absorption threshold is lower than the silica bulk. 

In both cases, the fine composition analysis indicates that \ch{Si3N4} ionized first. This is particularly visible in the top illumination case, when observing the interfaces between pristine nano-layers and modified ones. The \ch{Si3N4} nano-layers act as 'wicks' feeding the surrounding material with photo-dissociated, ionized nitrogen and silicon. This observation is expected as the band gap of \ch{Si3N4} is effectively lower than for \ch{SiO2} (typically around 5 eV for \ch{Si3N4} versus 9 eV for \ch{SiO2}) \cite{iqbalElectronicStructureSilicon1987,nikolaouInertAmbientAnnealing2015}. As \ch{Si3N4} has a significantly higher index, the existence of a guided mode in the confined layer is also to be expected in certain cases corresponding to the photonics bandgap formed by the layer structures (Bloch-wave). 
The samples are designed as quarter-wave stacks, therefore a lower modification threshold of the \ch{Si3N4} layers would be consistent with the modulation of the lasers' electrostatic field inside the multi-layers, which is more intense towards the higher-index layers, as reported in the literature even for layers of sub-micron dimensions \cite{csajbokFemtosecondDamageResistance2016,mangoteFemtosecondLaserDamage2012}. This would lead to the formation of a plasma early during the pulse-matter interaction, where the power contained in the pulse's "tail" is enough to reach the energy threshold for non-linear absorption in these precise locations. In the bottom exposure case, this can explain the presence of modifications both in the bulk and in the multi-layers stack, this effect being potentially enhanced by the beam elongation induced by the pulse's propagation through the substrate. 

In Fig.~\ref{experimental}a, the measured reflectance spectrum presents a short but noticeable peak around 1030 nm, equivalent to 18\% of reflectance. This means that multiple reflections of a fraction of the incoming pulses between the various layers are possible. This may explain the peculiar shape of the modifications visible in Fig.~\ref{TEM_1}, where the top layers, exposed to a greater amount of secondary reflections, are more likely to get damaged than the lower layers, and multiple reflections can partially explain the propagation of the modifications along the surface plane of the layers.

In the following summary, we draw a few hypothesis to explain the formation of these complex structures. This scenario remains phenomenological and would require further dynamical observations to confirm some of its key features. Keeping these reservations in mind, a possible sequence for both plasma formation and material modifications is as follows:

\begin{enumerate}
	\item Seed electrons are initially produced through multiphoton absorption in the \ch{Si3N4} layers due to the modulation of the laser’s electrostatic field by the Bragg reflector, as well as the lower ionization threshold of silicon nitride compared to silicon oxide. This is valid for both exposure conditions, top and bottom. However, in the case of bottom illumination, the Bragg reflector acts as a mirror for the effective laser wavelength as the beam propagates inside \ch{SiO2} instead of air. This would explain why in the case of bottom illumination, the first layers to be modified are always at the boundary between the Bragg reflector and the multilayer. These layers rapidly build up a plasma, through subsequent ionization processes, which is confined in the nano-layers through additional electron excitation mechanisms.
	\item As the confined layered-plasma structure density increases, the dielectric functions acquires a metallic behavior and favors the propagation of energy sideways, along the \ch{Si3N4}/\ch{SiO2} interfaces. It may share similarities with plasmon-polariton coupling reported at interfaces between layers supporting transverse optical Tamm states in the case of normal incidence exposure \cite{vetrovOpticalTammStates2017}.
	\item To support intermixing events, we propose the following mechanism. As the plasma builds up into a more homogeneous volume and in temperature, ions gain a sufficiently high energy to diffuse between multi-layers as well as in supporting substrate as noticed in the case of bottom illumination, mimicking an ion-implantation process. To support these observations, we note the presence of nitrogen down to a few microns within the substrate itself in the case of bottom illumination, where it was not present in first place, and the redistribution of oxygen throughout the modified zones for both exposure configurations. 
	\item As multiple pulses impact a same exposure zone at an average rate of 352 - 176 pulses $\mu m^{-2}$ with a time separation of 2 - 4 $\mu$s (depending on the repetition rate), pulse-to-pulse additive effects occur, and may account for some of the self-organized structures observed, just like in the case of nanogratings formation described previously in bulk silica \cite{shimotsumaSelfOrganizedNanogratingsGlass2003,bhardwajOpticallyProducedArrays2006a, liaoHighfidelityVisualizationFormation2015a, rudenkoRandomInhomogeneitiesPeriodic2016a}, possibly enhanced and stimulated by the presence of the Bragg reflector.  
	\item The fast intermixing and the ionic mobility produces the seeds for nanocrystalline silicon phase formation. In both cases, these nanocrystalline phases seem to have nucleated preferentially in the vicinity of interfaces. A possible explanation is that these are  regions of higher pressure confinement and higher defect density, and hence, higher phase nucleation probability. 
\end{enumerate}

\section{Conclusion}
This work investigates the interaction of ultrafast lasers with stacked \ch{SiO2}/\ch{Si3N4} sub-wavelength thick layer coatings, deposited on a fused silica substrate as a one-dimensional periodic structure. The exposure conditions lead to strong and controlled intermixing between layers, and to the formation of cubic-diamond silicon nanocrystals dispersed within the laser-affected zones. 

Two exposure cases are considered, leading to an asymmetric behavior related to the properties of the Bragg reflector. Top illumination, i.e. laser beam traveling through the multi-layers first, leads to laser-induced modifications initiated in the \ch{Si3N4} layers, propagating into the \ch{SiO2} layers and stretching laterally significantly more than the beam waist at the focus, unraveling a strong coupling between the incoming wave and a coupled mode in the \ch{Si3N4} higher density layers. Bottom illumination, i.e. laser beam focused in the substrate yields another outcome, is characterized by a more homogeneous volume where strong intermixing occurs and by the diffusion of atoms from the multi-layers into the substrate, where self-organization phenomena are observed as well, possibly enhanced by the presence of the Bragg reflector.

The exposure of multi-layer substrates with femtosecond lasers potentially offers a powerful method for the selective and confined formation of crystalline compounds in amorphous dielectric matrices, starting up from discrete elements of well-defined composition and benefiting from the intrinsic advantages of non-linear absorption.

\begin{acknowledgments}
The authors would like to thank the Swiss National Science Foundation (SNF) for funding this research through grant number CRSII5\_180232. Ruben Ricca would like to thank Simone Frasca for the help with the layers design, Dr. Julien Gateau for the feedback and the help with the laser machining, Dr. Lucie Navratilova for the preparation of the TEM lamellas, as well as Dr. Nicolas Descharmes and Dr. Raphaël Barbey for providing the specimens.
\end{acknowledgments}

\subsection*{Authors contributions}
YB and RR designed the experiments. RR performed the experiments, processed the data, prepared the figures. RR and YB discussed and interpreted the results of SEM and Raman observations. VB and RR performed the TEM observations and interpreted their results. RR and YB wrote the manuscript. All authors revised and discussed the article before submission. YB designed and supervised the project.

\subsection*{Disclosures}
The authors declare no conflicts of interest

\subsection*{Data Availability Statement}
The data that support the findings of this study are available from the corresponding author upon reasonable request.

%%%%%%%%%%%%%%%%%%%%%%% References %%%%%%%%%%%%%%%%%%%%%%%%%
\bibliography{Manuscript}

\end{document}